\documentclass[10pt,3p]{elsarticle}

\usepackage{epsfig,inputenc,amsfonts}
\usepackage{graphicx,amssymb,amsmath,color}
\definecolor{beige}{RGB}{245,245,220}
\definecolor{linen}{RGB}{250,240,230}
\definecolor{lightyellow}{RGB}{255,255,224}
\definecolor{thistle1}{RGB}{255,225,255}
\definecolor{MistyRose2}{RGB}{238,213,210}

\def\vrel{\vec v_g}
\def\vrelz{\vec v_{g,0}}
\def\tdt{t{+}\Delta t}

\journal{Journal of Computational Physics}

\begin{document}
\begin{frontmatter}

\title{An Adaptive Hierarchical Domain Decomposition Method for 
Parallel Contact Dynamics Simulations of Granular Materials}

\author{Zahra Shojaaee\corref{cor}}
\ead{zahra.shojaaee@uni-duisburg-essen.de}

\author{M.\ Reza Shaebani}

\author{Lothar Brendel}

\author{J\'anos T\"or\"ok}

\author{Dietrich E.\ Wolf}

\address{Computational and Statistical Physics Group, 
Department of Theoretical Physics, University of 
Duisburg-Essen, 47048 Duisburg, Germany}

\cortext[cor]{Corresponding author}
\begin{abstract}
  A fully parallel version of the Contact Dynamics (CD) method is
  presented in this paper. For large enough systems, $100\%$ efficiency
  has been demonstrated for up to $256$ processors using a
  hierarchical domain decomposition with dynamic load balancing. The
  iterative scheme to calculate the contact forces is left domain-wise
  sequential, with data exchange after each iteration step, which
  ensures its stability. The number of additional iterations required
  for convergence by the partially parallel updates at the domain
  boundaries becomes negligible with increasing number of particles,
  which allows for an effective parallelization. Compared to the
  sequential implementation, we found no influence of the
  parallelization on simulation results.
\end{abstract}

\end{frontmatter}

\section{Introduction}
\label{Introduction}

{\it Discrete element method} simulations have been widely employed 
in scientific studies and industrial applications to understand the 
behavior of complex many-particle systems such as granular materials. 
The common property of these methods is that the time evolution of 
the system is treated on the level of individual particles, i.e.\ 
the trajectory of each particle is calculated by integrating its 
equations of motion. Among the discrete element methods, {\it soft 
particle molecular dynamics} (MD) \cite{Cundall79,Luding04}, {\it 
event driven} (ED) \cite{Rapaport80,Walton86} and {\it contact 
dynamics} (CD) \cite{Jean92,Moreau94,Jean99,Brendel04} are often 
used for simulating granular media. 

{\em Molecular dynamics} is the most widely used algorithm for
discrete element simulations. For granular materials, the contact
forces between the soft particles stem from visco-elastic force
laws. Interactions are local, therefore efficient parallelization is
possible \cite{Plimpton95,Nyland97,Deng00,LAMMPS} with 100\%
efficiency for large systems (Throughout the paper, the performance of
a parallel algorithm is quantified by the usual quantities: the {\em
  speedup} $\mathcal{S}(N_p)$, which is the ratio of the run time of
the non-parallel version on a single processor to the run time of the
parallel version on $N_p$ processors, and the {\em efficiency} 
$\mathcal{E}{=}\mathcal{S}/N_p {\times} 100\%$). The time step and therefore the
speed of MD simulations is strongly limited by the stiffness of the
particles, as collisions must be sufficiently resolved in
time. Molecular dynamics is efficient for dense systems of soft
particles, but much less so for hard particles and
dilute systems.

The {\em event driven dynamics} \cite{Haff83,McNamara94} considers
particle interactions of negligible duration compared to the time
between collisions. Particle collisions are thus treated as
instantaneous events, and trajectories are analytically integrated in 
between. This makes ED favorable in dilute granular systems, where the 
above condition holds. The parallelization of this algorithm poses 
extreme difficulties, since the collisional events are taken from a 
{\em global} list, which in turn is changed by the actual collision. 
In general, a naive domain decomposition leads to causality problems. 
The algorithm presented in \cite{Miller03} conserves causality by 
reverting to an older state when violated. The best efficiency reached 
so far is a speedup proportional to the square root of the number of 
processors \cite{Miller03}.

In contrast to ED, lasting contacts between rigid bodies are
considered in the realm of \emph{(multi)-rigid-body dynamics}. Common
to all its realizations is the treatment of contact forces as
constraint forces, preventing interpenetration and, to a certain
extent in the case of frictional contacts, sliding. When applying the
rigid body modelling to problems like e.g.\ robotics
\cite{Lotstedt82,Stewart_Trinkl00} or granular media
\cite{CDvalidation1,CDvalidation2,CDvalidation3,CDvalidation4},
different algorithms can in principle be used. Approximations with
respect to the constraint of dry Coulomb friction enable the usage of
powerful standard techniques for \emph{linear complementary problems}
(LCP) \cite{Preclik_etal_09}. Other algorithms keep the isotropic
friction cone, using a solver based on a modified time stepping scheme
leading to a \emph{cone complementary problem} (CCP) for the
simulation of frictional contact dynamics \cite{Anitescu_06}. Other
approximations, leading to \emph{fast frictional dynamics} (FFD)
\cite{Kaufman_etal_05}, yield a computational cost being only linear
in the number of contacts and thus allow for impressive system sizes
in terms of the number of particles. For investigations of e.g.\ the
stress field in granular media, these approximations are prohibitive,
though, and thus the \emph{non-smooth contact dynamics} method
\cite{Jean99}, or commonly just \emph{contact dynamics}, is widely
employed. We will sketch the principle of this iterative procedure in
section \ref{CD-algorithm}. Parallelization of the FFD method is
straightforward and efficient \cite{Iglberger_09,Iglberger_10}, on the
other hand, the parallel version suffers also from the undesired
approximations. The parallel implementation of the CCP algorithm by
the use of the Graphics Processing Unit (GPU) for large-scale
multibody dynamics simulations is presented in
\cite{Tasora_etal_11}. In the present work we investigate the impact
of the parallelization on the numerical solution of the CD method
going beyond \cite{Iglberger_09,Iglberger_10,Tasora_etal_11}.

Providing a parallel CD code is motivated by the need for large-scale
simulations of dense granular systems of hard particles. The
computation time even scales as ${\cal O}(N^{1+2/d})$ with the number
of particles in CD \cite{Brendel04} ($d$ is the dimension of the
system), while it grows linearly with $N$ in MD. However,
parallelization of CD poses difficulties as in general the most time
consuming part of the algorithm is a global iteration procedure, which
cannot be performed completely in parallel. So far, a static
geometrical domain decomposition method has been proposed in
Ref.~\cite{Breitkopf99}, and a partially parallel version is
introduced in Ref.~\cite{Renouf03}, where only the iterative solver is
distributed between shared memory CPUs. In the former work, the force
calculation is studied just on $8$ processors and in the
latter, already with $16$ processors the performance efficiency is
below $70\%$. None of these studies
deal with computational load balancing during the execution of the
code.

There is a large variety of domain decomposition methods proposed for
parallel particle simulations in the literature, from {\it Voronoi
tessellation} \cite{Zhakhovskii04} to {\it orthogonal recursive
bisection} (ORB) \cite{Salmon90,Warren93}. For the parallelization of
CD the size of the interfaces between domains is more crucial than for
MD, since besides communication overhead it also influences the
parallel/sequential nature of the global iteration. So the ORB methods
are the most suited for the CD code together with adaptive load
balancing approaches \cite{Fleissner08}, which is not only important in
heterogeneous clusters but also in the case of changing simulation
setup and local particle/contact density.

In the present work, we introduce a completely parallel version of the
contact dynamics method using MPI communication with orthogonal
recursive bisection domain decomposition for an arbitrary number of
processors. The method minimizes the computational costs by optimizing
the surface-to-volume ratio of the subdomains, and it is coupled with
an adaptive load balancing procedure. The validation of our code is
done by numerical simulations of different test systems. We presented 
our implementation in two dimensions and for spherical particles. 
However, our code is also capable of handling polygonal particles 
and the extension to three dimensions is straightforward.

This article is organized as follows. The contact dynamics method is 
described briefly in Sec.~\ref{CD-Method} and particular attention is 
paid to the numerical stability of the sequential and parallel update 
schemes, and to the identification of the most time consuming parts 
of the code. In Sec.~\ref{ParallelCD} we present an adaptive 
domain decomposition method, and implement it in a parallel version 
of the CD algorithm. The results of some test simulations with respect 
to the performance of the parallel CD code are presented in 
Sec.~\ref{NumericalResults} and the effect of our parallelization 
approach on the physical properties of the solutions are investigated. 
We conclude the paper in Sec.~\ref{conclusions} with a summary of the 
work and a brief discussion.

\section{Contact Dynamics Method}
\label{CD-Method}

\subsection{A brief description of the CD algorithm}
\label{CD-algorithm}

In this section, we present the basic principles of the CD
method \cite{Jean99} in a language closer to MD and
with special emphasis on those parts where changes are applied in the
parallel version of the code. For more details and a broader overview
cf.\ \cite{Jean99,Stewart_00}. The central point is that the forces are not calculated 
from particle deformation, instead they are obtained from the 
constraints of impenetrability and friction laws. Imposing 
constraints requires implicit forces, which are calculated to 
counteract all movement that would cause constraint violation.

In general for molecular dynamics, where trajectories are smooth (soft
particle model), simulation codes use second (or higher) order schemes 
to integrate the particle positions. In CD method, the non-smooth
mechanics (hard particle limit) requires strong discontinuity, which 
can only be achieved by first order schemes. Thus we apply a {\it 
first-order Euler} scheme for the time stepping of particle $i$:
\begin{equation}
  \vec v_i(\tdt)= \vec v_i(t) +
  \frac{1}{m_i} \vec F_i \, \Delta t,
  \label{veloc-update}
\end{equation}
\begin{equation}
  \vec r_i(\tdt) = \vec r_i(t) +
  \vec v_i(\tdt)\Delta t \, ,
  \label{pos-update}
\end{equation}
which determines the new velocity $\vec v_i$ and position 
$\vec r_i$ of the center of mass of the particle after a time step 
$\Delta t$. The effective force on particle $i$ is denoted by $F_i$. 
The scheme is semi-implicit in the sense that the right-hand-side
velocities are (necessarily) the ones at time $\tdt$ while forces
other than the constraint forces may be treated implicitly or
explicitly. The size of the time step $\Delta t$ is chosen such that the relative
displacement of the neighboring particles during one time step is much
smaller compared to the size of particles and to the radius of
curvature of contacting surfaces. Similar equations are used for the
rotational degrees of freedom, i.e.\ to obtain the new angular
velocity $\vec \omega_i(\tdt)$ (caused by the new torque $\vec
T_i(\tdt)$), and the new orientation of particle $i$.

For simplicity, in the following we assume that particles are dry and
non-cohesive having only unilateral repulsive contact forces.
Furthermore, we assume perfectly inelastic particles, which remain in
contact after collision and do not rebounce. The implicit scheme must
fulfill the following two constraints:
\begin{enumerate}
 \item[(i)] the {\it impenetrability} condition: the 
overlapping of two adjacent particles has to be prevented by 
the contact force between them.
\item[(ii)] the {\it no-slip} 
condition: the contact force has to keep the contact from sliding 
below the Coulomb friction limit, i.e.\ the tangential component of the
contact force cannot be larger than the friction coefficient times the
normal force.
\end{enumerate}
The contact forces should be calculated in such a way that the
constraint conditions are satisfied at time $\tdt$, for the new
particle configuration \cite{Brendel04}. Once the total force and
torque acting on the particles, including the external forces and also
the contact forces from the adjacent particles, are determined, one can
let the system evolve from time $t$ to $t{+}\Delta t$.

\begin{figure}[b]
\begin{center}
\includegraphics*[scale=0.65]{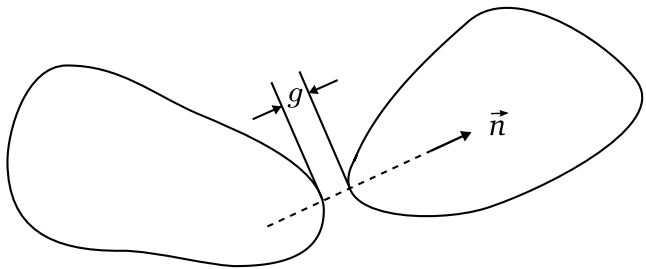}
\end{center}
\caption{Schematic picture showing two adjacent rigid particles.}
\label{Fig-1}
\end{figure}

Let us now consider a pair of neighboring rigid particles in
contact or with a small gap between them as shown in 
Fig.~\ref{Fig-1}. We define $\vec n$ as the unit vector
along the shortest path of length $g$ between the surfaces of the two
particles. The relative velocity of the closest points is called the
{\it relative velocity of the contact} $\vrel$. In the
case that the particles are in contact, the gap $g$ equals to zero,
and $\vec n$ denotes the contact normal.

We first assume that there will be no interaction between the two
particles at $t{+}\Delta t$, i.e.\ the new contact force $\vec
R(t{+}\Delta t)$ equals to zero. This allows the calculation of a
hypothetical new relative velocity of the two particles $\vrelz(\tdt)$
through Eq.~(\ref{veloc-update}), only affected by the remaining
forces on the two particles. The new gap reads as:
\begin{equation}
g(t{+}\Delta t)=g(t)+\vrelz
(t{+}\Delta t) {\cdot} {\vec n} \Delta t.
  \label{new-gap}
\end{equation}
If the new gap stays indeed positive ($0{<}g(t{+}\Delta t)$) then no 
contact is formed and the zero contact force is kept: $\vec R(t+\Delta
t){=}0$. 

On the other hand, if the gap turns out to be negative ($g(t{+}\Delta t)
\leq 0$), a finite contact force must be applied. First, we determine
the new relative velocity from the condition that the
particles remain in contact after the collision, 
\begin{equation}
0\equiv g(\tdt){\vec n} = g(t){\vec n}+\vrel (\tdt) \Delta t
  \label{no-overlap-condition}
\end{equation}
Here we assume sticking contacts with no relative velocity in the 
tangential direction ($\vrel^{\,t} (\tdt){=}0$), which implies that the 
Coulomb condition holds. The new contact force satisfying the 
impenetrability can be obtained using Eq.~(\ref{veloc-update}) as
\begin{equation}
\vec R(t{+}\Delta t) = \frac{\bf M}{\Delta t}
  \biggl(\vrel (t{+}\Delta t)-\vrelz (t{+}\Delta t)\biggr)=
\frac{ - {\bf M}}{\Delta t}
  \left(\frac{g(t)}{\Delta t}\vec n + \vrelz (t{+}\Delta t)\right)
  \label{new-force-1}
\end{equation}
where the mass matrix $\bf M$, which is built up from the masses and 
moments of inertia of both particles \cite{Brendel04}, reflects 
the inertia of the particle pair in the sense that ${\bf M}^{-1}\vec R$ corresponds 
to the relative acceleration of the contacting surfaces induced 
by the contact force $\vec R$.

At this point, we have to check for the second constraint: the
Coulomb friction. Let us first define the normal and tangential
contact forces:
\begin{eqnarray}
  \label{tangential-normal-contact-force}
  R_n(t) &\equiv& \vec R(t) {\cdot} \vec n \;\; , \cr
  \vec R_t(t) &\equiv& \vec R(t)- R_n(t) \vec n \;\; .
\end{eqnarray}
Then the Coulomb inequality reads as
\begin{equation}
\left| \vec R_t (t{+}\Delta t) \right| \le \mu R_n (t{+}\Delta t)\,,
  \label{Coulomb-cone}
\end{equation}
where $\mu$ is the friction coefficient (being the same 
for static and dynamic friction, the standard Coulomb model of dry 
friction \cite{Stewart_00}). If the inequality (\ref{Coulomb-cone}) 
holds true, then we have already got the correct contact forces. 
Otherwise, the contact is sliding, i.e.\ $\vrel (t{+}\Delta t)$ has 
a tangential component and Eq.~(\ref{no-overlap-condition}) reads
\begin{equation}
0\equiv g(\tdt) = g(t)+{\vec n} {\cdot} \vrel (\tdt)  \Delta t \,,
\label{no-overlap-condition-new}
\end{equation}
which determines the normal component of $\vrel (t{+}\Delta t)$. The 
remaining five unknowns, three components of the contact force
$\vec R(\tdt)$ and two tangential components of the relative velocity, 
are determined by the following two equations: 

(i) Impenetrability by
combining Eqs.~(\ref{no-overlap-condition}) and 
(\ref{new-force-1})
\begin{equation}
\vec R(\tdt) {=} \frac{\bf M}{\Delta t} 
  \biggl(-\frac{g(t)}{\Delta t}\vec n + \vrel^{\,t}
(t{+}\Delta t) - \vrelz (t{+}\Delta t)\biggr).
  \label{new-force-3}
\end{equation}

(ii) Coulomb condition
\begin{equation}
  \vec R_t (t{+}\Delta t)=-\mu R_n (t{+}\Delta t)
  \frac{\vrel^{\,t} (t{+}\Delta t)}
  {\left|\vrel^{\,t} (t{+}\Delta t) \right|}.
  \label{tangential-force}
\end{equation}
In two dimensions and for spheres in three dimensions, 
these equations have an explicit analytical solution, otherwise one 
has to resort to a numerical one\cite{Jean99}.

\begin{figure}[t]
\centering
\includegraphics*[trim=3cm 20.3cm 5cm 4cm,scale=0.99]{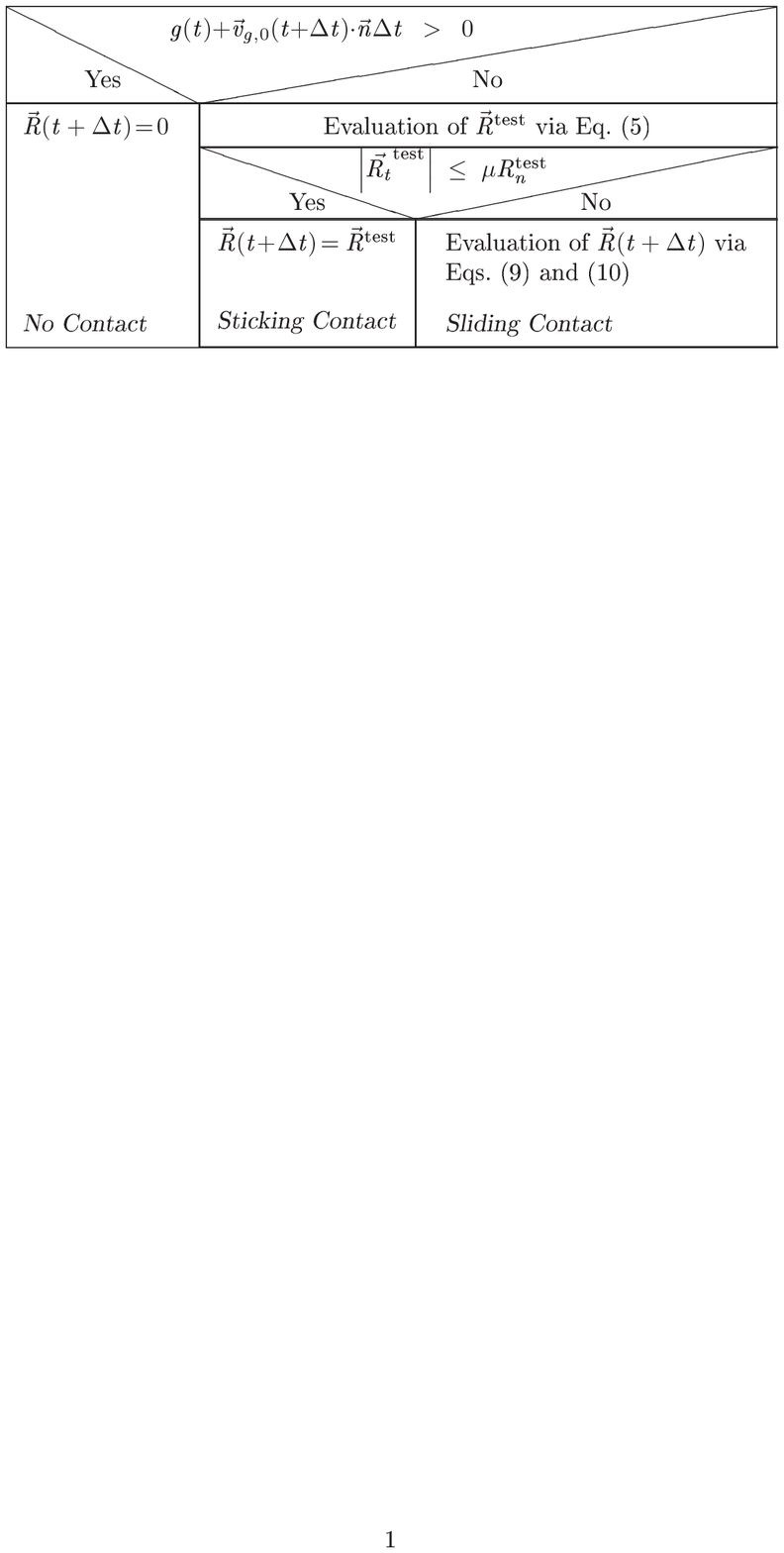}
\caption{The force calculation process for a single contact.}
\label{Fig-2}
\end{figure}

Figure~\ref{Fig-2} summarizes the force calculation process for a single 
incipient or existing contact. Assuming that all other forces acting on 
the participating particles are known, the Nassi-Shneiderman diagram 
\cite{Nassi73} in Fig.~\ref{Fig-2} enables us to determine the contact 
force.

The above process assumes that apart from the contact forces all other
interactions are known for the selected two particles. However, in
dense granular media, many particles interact simultaneously and form
a contact network, which may even span the whole system. In such cases,
the contact forces cannot be determined locally because each unknown
contact force depends on the adjacent unknown contact forces acting on
the particles. In order to find the unilateral frictional forces
throughout the entire contact network, an {\it iterative} method is mostly
used in CD as follows: At each iteration step, we choose the contacts
randomly one by one and calculate the new contact force considering the 
surrounding contact forces to be already the correct ones. It is 
natural to update the contact forces sequentially in the sense that 
each freshly calculated force is immediately used for further force 
calculations. One iteration step does not provide a globally consistent 
solution, but slightly approaches it. Therefore, the iteration has to 
be repeated many times until the forces relax towards an admissible state.
To assess whether or not the convergence is achieved, we measure the 
relative change of each contact force $\vec R_i$ at each iteration 
step $j$, as well as the relative change in the average contact force 
$\vec R_\text{avg}$ at this iteration step. Generally, we choose one 
of the following convergence criteria to stop the force calculation 
procedure: 
\begin{enumerate}
\item[(I)] \emph{local convergence test}: if, at least for $90\%$ of 
the contacts, the following condition holds
\begin{equation}
\displaystyle\frac{(\vec R_i^{^j}{-} \vec R_i^{^{j-1}})^2}{(\vec R_i^{^j} 
{+} \vec R_i^{^{j-1}})^2}\;{<}\;\alpha, \nonumber \\
\end{equation}
and the rest of contacts fulfill
\begin{equation}
(\vec R_i^{^j}{-} \vec R_i^{^{j-1}})^2 \; {<}\; \alpha (\vec R_\text{avg}^{^{j-1}})^2. \nonumber \\
\end{equation}
\item[(II)] \emph{global convergence test}: if the relative change 
in the average contact force falls below the threshold value $\alpha$, i.e.
\begin{equation}
\displaystyle\frac{(\vec R_\text{avg}^{^j}{-} \vec R_\text{avg}^{^{j-1}})^2} 
{(\vec R_\text{avg}^{^j} {+} \vec R_\text{avg}^{^{j-1}})^2}\;{<}\;\alpha.
\nonumber \\
\end{equation}
\end{enumerate}
We have chosen $\alpha{=}10^{-6}$ in all simulations.

The precision of the solution increases smoothly with the
number of iterations $N_I$, with the exact solution being only reached for
$N_I\to\infty$. Of course we stop at finite $N_I$. It is optional to
use a fixed number of iterations at each time step, or to prescribe a
given precision to the contact force convergence and let $N_I$ to vary
at each time step.

Breaking the iteration loop after finite iteration steps is an
inevitable source of numerical error in contact dynamics simulations,
which mainly results in overlap of the particles and in spurious
elastic behavior \cite{Unger02}. Occurring oscillations are a sign 
that the iterations were not run long enough to allow the force 
information appearing on one side of the system to reach the other 
side. This effect should be avoided and the number of iterations 
should be chosen correspondingly \cite{Unger02}.

\begin{figure}[t]
\begin{center}
\includegraphics*[width=0.50\textwidth]
{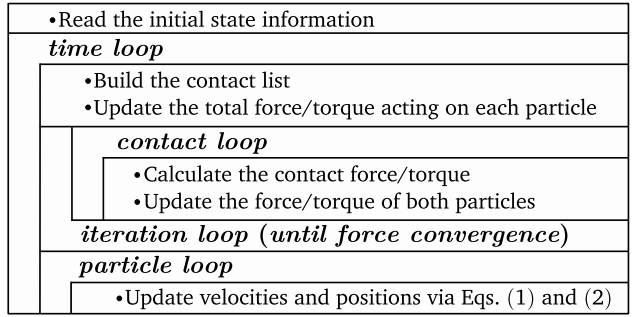}
\end{center}
\caption{The diagram of the main steps of the contact 
dynamics algorithm.}
\label{Fig-3}
\end{figure}

Once the iteration is stopped, one has to update the particle positions
based on the freshly calculated forces acting on the particles.
Figure~\ref{Fig-3} concludes this section with a diagram
depicting the basic steps of the contact dynamics algorithm.

The question of successful convergence in general 
is difficult (cf.\ \cite{Stewart_98,Jourdan_Alart_Jean_98}) but in 
practice convergence turns out to be given and hence the CD method 
has been experimentally validated in different instances, e.g.\ as 
in calculating the normal contact force distribution in static 2D 
and 3D granular packings especially for weak forces that are 
experimentally difficult to access \cite{CDvalidation1}, 
investigating the dynamics of granular flows e.g.\ monitoring the 
evolution of the contact orientations and shear band formation in a 
biaxial shear cell \cite{CDvalidation2}, studying the mechanical 
properties of cohesive powders \cite{CDvalidation3}, and predicting 
the refraction of shear zone in layered granular media 
\cite{CDvalidation4}.

\subsection{CPU time analysis}
\label{CPUTimeAnalysis}

The CD algorithm described in the previous section has three main
parts: (i) The contact detection, (ii) the force calculation
(iteration), (iii) the time evolution. In this section we analyze the
CPU consumption of all these parts.

Given a system and the contact detection algorithm, the time
consumption of parts (i) and (iii) can be easily estimated. On the
other hand, the computational resource needed by part (ii) is strongly
influenced by the number of iterations. If one uses extremely high
values of $N_I$, part (ii) will dominate the CPU usage. This led Renouf
{\it et al.} \cite{Renouf03} to the conclusion that parallelizing the
force calculation is enough.

Our view is that the situation is more delicate and it is demonstrated
by a simulation in which diluted granular material is compressed until
a dense packing is reached \cite{Shaebani09}. The system consists of
$1000$ polydisperse disks in two dimensions with friction coefficient
$\mu{=}0.5$. The stopping criteria for the iteration was the
fulfillment of any of the two conditions:
\begin{enumerate}
\item[(1)] The global convergence criterion is fulfilled 
(see Sec.~\ref{CD-algorithm} for details).
\item[(2)] $N_I\ge 200$
\end{enumerate}

\begin{figure}
\begin{center}
\includegraphics*[width=0.70\textwidth, trim=0cm 0cm 0cm 0cm]
{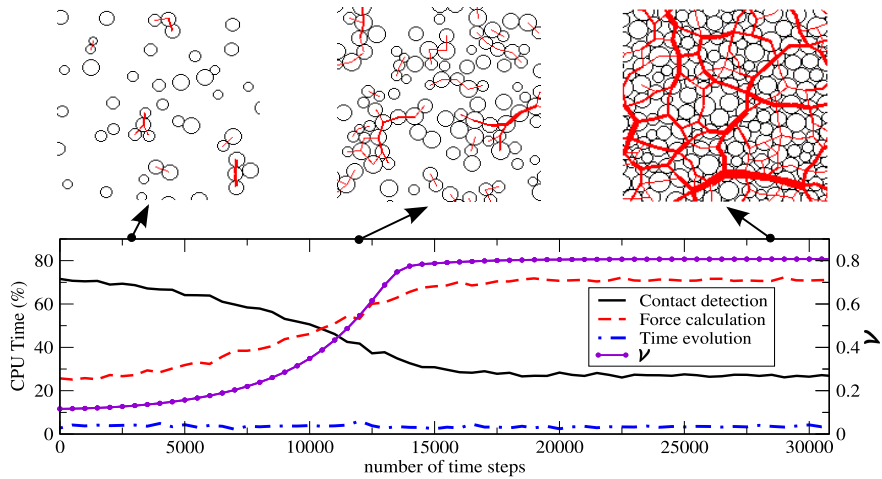}
\end{center}
\caption{(color online) The percentage of CPU time consumption (lines) 
and the packing fraction $\nu$ (purple line, full circles) as a function 
of time. The insets show typical configurations of particles at different 
packing fractions. The thickness of the inter-center connecting red lines 
is proportional to the magnitude of the contact force.}
\label{Fig-4}
\end{figure}

Figure~\ref{Fig-4} shows the evolution of the relative CPU time
consumption of the different parts of the algorithm. The time stepping
contribution always remains less than $5\%$, and the rest is divided
between the other two subroutines. Initially, the contact detection
task consumes the majority of the computational time. After a while,
clusters of contacting particles form, and the cost of force
calculation increases and the iterative solver gradually becomes the
most time consuming part of the code. Note that the contribution of
the solver saturates to $70\%$ of the total elapsed time. If only the
force calculation part is executed in parallel, even with
$\mathcal{E}_\text{force}=100\%$, the remaining $30\%$ non-parallel
portions set an upper limit to the overall efficiency $\mathcal{E}$
and the speedup $\mathcal{S}$ of the code (${\mathcal{E}}_\text{max}
\approx 80\% $ and ${\mathcal{S}}_\text{max}\approx 4$). Therefore, we
aim to provide a fully parallel version of CD which operates
efficiently in all density regimes.

\subsection{Sequential versus parallel update scheme}
\label{SeqvsParallel}

As we pointed out in Sec.~\ref{CD-algorithm}, the problem of finding the 
unilateral frictional contact forces that satisfy the constraint conditions 
cannot be solved locally in a dense granular system. In order to evaluate 
the new value of a single contact, one has to know the new values of the 
adjacent contact forces, which are unknown as well, i.e.\ all contact 
forces are coupled in a cluster of contacting particles. Note that this 
is a consequence of the infinite stiffness of the particles; a single 
collision influences the entire network of contact forces between 
perfectly rigid particles. This problem is solved by iterating through
all contacts many times until a given precision is reached.

Similarly to solving the Laplace equation, the information about a
disturbance (e.g.\ collision of a new particle) appearing on one side
of a cluster must diffuse through the whole cluster to satisfy the
constraints. Actually, the iteration scheme is very similar to two
traditional schemes for solving a set of linear equations 
\cite{NumericalRecipes}, albeit with nonlinearities introduced by the
change of contact states (repulsive vs.\ force-less, sticking vs.\
sliding): the Jacobi scheme and the Gauss-Seidel scheme, corresponding 
to parallel and sequential contact updating, respectively. 

\begin{figure}[t]
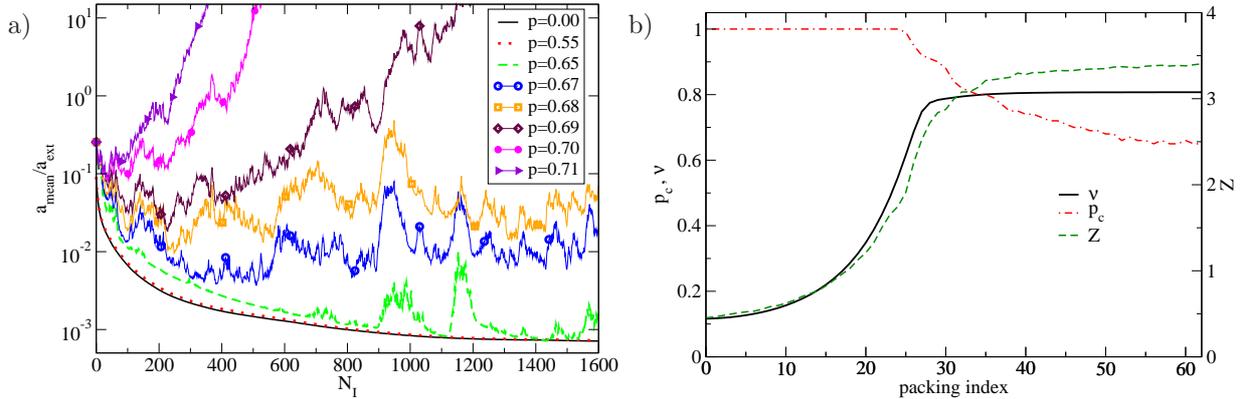

\begin{center}
\raisebox{32ex}{a)}\includegraphics*[width=0.47\textwidth, trim=0cm 0cm 0cm 0cm]
{Figure-5-a.eps}
\raisebox{32ex}{b)}\includegraphics*[width=0.465\textwidth, trim=0cm 0cm 0cm 0cm]
{Figure-5-b.eps}
\end{center}
\caption{(color online) (a) The mean acceleration of the particles 
$a_\text{mean}$ scaled by $a_\text{ext}{=}2{\bar r}P_\text{ext}/{\bar m}$ 
(where $\bar r$ and $\bar m$ are the mean particle radius and mass, 
respectively, and $P_\text{ext}$ is the external pressure) in terms 
of the number of iterations $N_I$ for several values of the
``parallelness'' $p$ (cf.\ text). These results belong to 
the dense packing in the right panel of Fig.~\ref{Fig-4}. (b) The critical 
parallelness ratio $p_c$, the average coordination number $Z$, and the packing 
fraction $\nu$ for several configurations obtained during the time evolution 
of the system in the simulation presented in Fig.~\ref{Fig-4}.}
\label{Fig-5}
\end{figure}

Here we denote (i) {\em sequential}, where the contacts are solved one
by one using always the newest information available, which is a
mixture of new and old values, (ii) {\em parallel}, where all contacts 
are updated using the old values, and substituted with the new ones at 
the end of the iteration step. Needless to say that the second case is 
favored for parallel applications but instabilities may appear (like 
when combining the Jacobi scheme with Successive Over-Relaxation 
\cite{NumericalRecipes}). To study its impact, we investigated a mixed 
method, where a fraction $p$ of the contacts are updated in parallel 
and the rest sequentially. First, a static homogeneous packing is 
generated by applying an external confining pressure \cite{Shaebani09}. 
Next, the inter-particle forces are set to zero, while the positions of 
the particles and the boundary conditions are kept fixed. Now the code 
recalculates the contact forces within one time step with an 
unconstrained number of iterations until the convergence is reached. 
We check how many iteration steps are needed to find a consistent 
equilibrium solution with a given accuracy threshold. The results are 
shown in Fig.~\ref{Fig-5}(a).

It turns out that, on average, the number of iterations $N_I$ to reach 
a given accuracy level increases with increasing $p$. For high values 
of $p$, fluctuations appear and beyond $p_c \approx 0.65$ the iterative 
solver is practically unable to find a consistent solution. We discuss 
the consequence of this behavior for the parallel version of CD in 
Secs.~\ref{ParallelCD} and \ref{NumericalResults}.

In order to investigate the dependence of $p_c$ on the 
properties of the contact network, we take snapshots of the structure 
during the time evolution of the system in the simulation presented in 
Fig.~\ref{Fig-4}. The same procedure as mentioned above is then applied 
to each of these samples to obtain $p_c$. The results are shown in 
Fig.~\ref{Fig-5}(b). In dilute systems, the contacts form small isolated 
islands and the resulting set of equations is decomposed into smaller 
independent sets, so that even a completely parallel update scheme 
($p_c{=}1.0$) can be tolerated. However, the contact network in dense 
systems forms a set of fully coupled nonlinear equations which converges 
only if the parallelness factor $p$ is less than $p_c{\sim}0.65$. By varying 
the system size and the friction coefficient, we conclude that $p_c$ is 
mainly influenced by the degree of coupling between the equations which 
is reflected in the connectivity of the sample $Z$ [see Fig.~\ref{Fig-5}(b)].

Thus, the results of our numerical simulations reveal that the
sequential update scheme is quite robust and the force convergence is
reached smoothly, while the fully parallel update scheme is highly
unstable in dense systems. However, there is a limit of parallel update 
for which the iteration remains stable. This is important because the 
domain decomposition method allows for a sequential update only in the 
bulk of each domain, while the boundary contacts are updated in a parallel
way (cf.\ section \ref{ParallelAlgorithm}). This analysis suggests
that the ratio of bulk contacts to boundary ones after the
decomposition should never fall below 1. Fortunately, this is assured
in a domain decomposition context anyway.

\section{A parallel version of the CD algorithm}
\label{ParallelCD}

\subsection{The parallel algorithm}
\label{ParallelAlgorithm}

A parallel version of the CD algorithm based on the decomposition 
of the simulation domain is introduced in this section. The main 
challenge is to properly evaluate the inter-particle forces when 
the contact network is broken into several subnetworks assigned to 
different processors. The parallelization presented in this section is
valid only for spherical particles (disks in 2D), but it is
straightforward to extend it for other shape types.

At the beginning of the simulation, a domain decomposition function is
called to divide the system between $N_p$ processors. Regarding the
fact that neither the performance of the computing environment nor the
density distribution and the internal dynamics of the system are known
initially, a uniform distribution for all relevant factors is assumed
and initially the simulation domain is geometrically divided into
$N_p$ parts with the same volume. The details of the hierarchical
decomposition method are explained in Sec.~\ref{DomainDecompos}.

\begin{figure}[t]
\centering
\includegraphics*[width=0.70\textwidth, trim=0cm 0.0cm 0cm 0cm]
{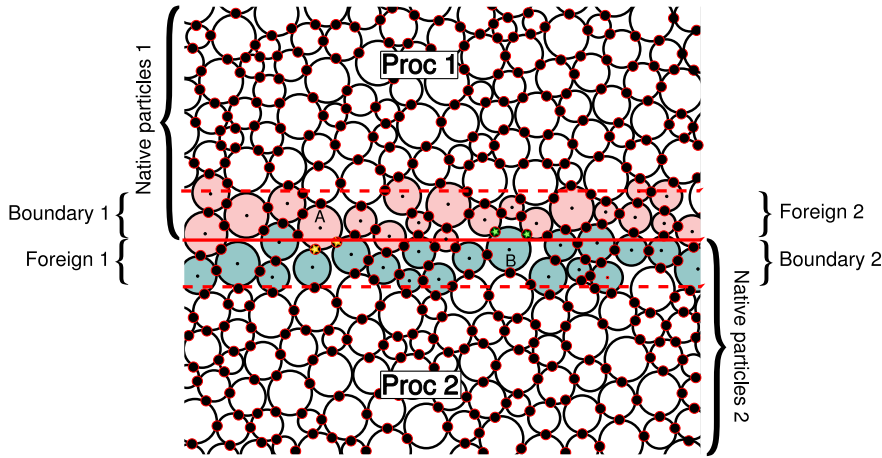}
\caption{(color online) Schematic picture showing two neighboring 
processors at their common interface. Their respective domain and boundary 
regions are marked. Particle A is a {\em native} 
particle of processor $1$ and is in contact (asterisks) with two 
{\em foreign} particles, namely {\em boundary} particles of 
processor $2$. The contacts are \emph{boundary} contacts of 
processor $2$ and thus \emph{foreign} ones to processor $1$. 
Particle B is a {\em boundary} particle of processor $2$ and 
has two contacts (asterisks) located inside the domain of 
processor $1$, i.e.\ they belong to the latter's \emph{boundary} contacts.}
\label{Fig-6}
\end{figure}

After establishing the domains, the particles are distributed among the
processors. Each processor maintains its set of {\em native} particles,
the center of mass of which lie within its domain. The next task is to
identify in each domain the \textit{boundary} particles, i.e.\ those
particles which may be in contact with
particles in other domains, as this information should be passed to
the neighbors. Two particles may come into contact if the gap is
smaller than $2 v_\mathrm{max}\Delta t$, where $v_\mathrm{max}$
is the maximum velocity in the whole system. So the maximal distance 
between the centers of mass of two particles, which may come into 
contact is
\begin{equation}
d\leq 2r_\mathrm{max}+2v_\mathrm{max}\Delta t,
\label{Eq:Bondary_width}
\end{equation}
where $r_\mathrm{max}$ is the radius of the largest particles. This
distance also defines the width of the \emph{boundary region} in which
particles may have contact with particles outside a processor's domain,
see also Fig.~\ref{Fig-6}.

While $r_\mathrm{max}$ is constant during the simulation,
$v_\mathrm{max}$ varies in time and space. For reasons described in
Sec.~\ref{DomainDecompos}, we use a global upper limit $\ell$ for the 
boundary size, which is unchanged during the whole simulation. It was 
explained in Sec.~\ref{CD-algorithm}, that the displacement of the 
particles must be small compared to particle size for contact dynamics 
to be valid. Therefore it is legitimate to define the upper limit for 
the particle displacement to be $0.1 r_\mathrm{max}$ and thus use the 
boundary size
\begin{equation}
\ell=2.2r_\mathrm{max} \,.
\end{equation}
Hence, a small amount of in principle irrelevant neighboring 
information is transferred. This is dominated by other effects, though, 
as will be shown in Sec.~\ref{DomainDecompos}.

After the identification of the \textit{boundary} particles, their
relevant data is sent to the corresponding neighbor processors, which
keep the information of these (to them) \textit{foreign}
particles. Since sender and receiver will always agree about the
forces acting on these particles, the receiver can evolve their state
on its own.

The next step is to identify actual and possible contacts between both
{\em native} and {\em foreign} particles. A position is assigned to
each contact, which is the middle of the gap (see
Fig.~\ref{Fig-1}). Obviously, for particles in touch, 
this is the contact point. Each processor builds a list of
\emph{native} contacts for the iteration loop exclusively from
contacts lying in its domain. The remaining ones are called
\emph{foreign} contacts and are in turn \emph{boundary} contacts of 
neighboring processors. During an iteration sweep, they will not be
updated but their forces enter the force calculation algorithm. Only
at the end of the sweep, each processor sends the new forces of its
\emph{boundary} contacts to its corresponding neighbor. This means 
that during an iteration sweep, foreign contacts always have the 
values from the last iteration, while native contacts are gradually
updated realizing a mixture of parallel and sequential update.

The convergence of the force calculation has to be checked after each
iteration sweep. This should be a global test, since the convergence in
different subdomains may be achieved at different iteration steps.
This task can only be completed by a single processor. Therefore, the
necessary data is collected and submitted to the root processor, which
makes a decision whether the iteration should continue or the
convergence is good enough and time stepping can take place. If 
further iterations are necessary, then only boundary \emph{contact}
information are exchanged among neighbors, as particles do not move
within the iteration loop. With new foreign contact values, each
processor can perform the next iteration sweep. If the iteration 
loop has finished, the particles are displaced according to the 
implicit Euler scheme of Eqs.~(\ref{veloc-update}) and 
(\ref{pos-update}). Every processor is responsible for its own native
particles (but evolves its copies of foreign particles as well).

Before starting the next time step, we have to
take care of the load balancing: Every processor broadcasts its own 
elapsed CPU time, which provides the required information to run the
load balancing function. The detailed description of this function is
presented in Sec.~\ref{LoadBalancing}.  If the load balancing function
redivides the simulation box, then each processor has to compare its
own particle positions to the new domain coordinates of all other
processors to determine to which processor each particle has to be
sent. This re-association of particles takes place also without domain
redivision as particles change domains simply due to their dynamics.

\begin{figure}[t]
\centering
\includegraphics*[width=0.62\textwidth]
{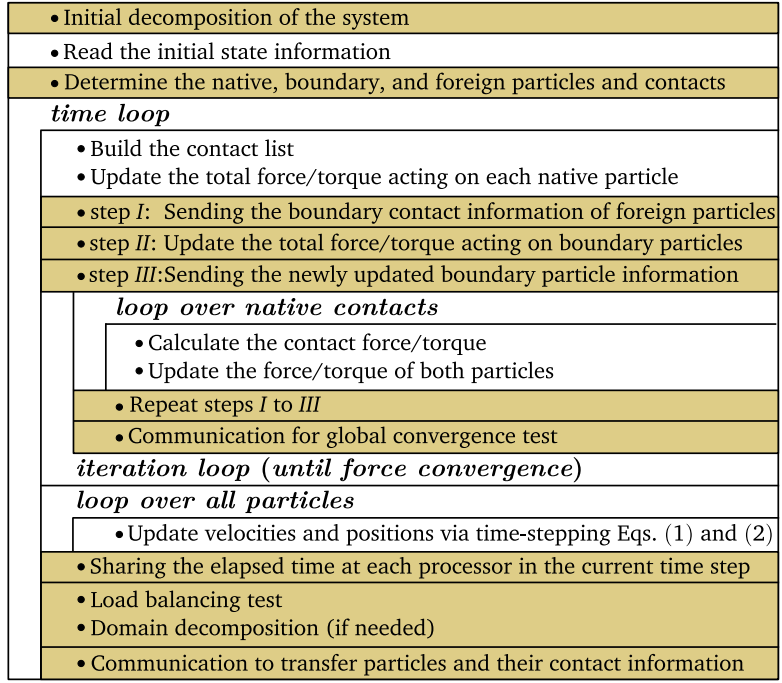}
\caption{(color online) The diagram of the parallel version of CD. 
The colored regions correspond to the new parts compared to the 
original CD algorithm shown in Fig.~\ref{Fig-3}.}
\label{Fig-7}
\end{figure}

Figure \ref{Fig-7} summarizes the parallel
algorithm. The main differences (highlighted in the diagram) are that
(i) at certain points data must be sent or received to neighboring
domains; (ii) the iteration scheme updates only native contacts
gradually, while foreign contacts are refreshed only after a complete
iteration sweep; (iii) load balancing and domain redivision checks take
place at the end of the time step.

A mixture of the sequential and the parallel update scheme occurs for
a fraction of the contacts. This fraction depends on the
surface-to-volume ratio of the subdomain. As discussed in
Sec.~\ref{SeqvsParallel}, a mixed update can become unstable if the
contribution of the parallel update exceeds a threshold of order
unity.  This limitation coincides with the standard limitation of
parallel computation that the boundary region should be negligible
compared to the bulk. In this sense, for reasonably large systems,
we do not expect any instability impact due to the parallel
update. Nevertheless, this issue is investigated in
Sec.~\ref{Sec-NI-NP}.

In the next section we introduce a hierarchical domain decomposition
method, which finds the best way to arrange the
orientation and location of the interfaces so that the
surface-to-volume ratio is minimal for a given number of processors.

\subsection{Hierarchical domain decomposition}
\label{DomainDecompos}

Before describing the domain decomposition, we have to investigate the
contact detection. This process, for which the brute force algorithm
scales as ${\cal O}(N^2)$ with the number of particles, can be
realized for different levels of polydispersity
\cite{Allen87,Wackenhut05,Ogarko10} within ${\cal O}(N)$
CPU cycles. We chose to implement the most widespread one, the
cell method \cite{Allen87}, which works well for moderate
polydispersity and which is the most suitable for parallel
implementation.

The cell method puts a rectangular grid of mesh size $a_x\times a_y$
on the simulation space. Each particle is assigned to its cell
according to its position, and the mesh size is chosen such that the
particles can only have a contact with particles from neighboring
cells and their own. That means, the cell diameter has essentially the
same meaning as the width of the boundary region $\ell$ and thus they
should coincide. On the other hand, the values $a_x$ and $a_y$ have to
be chosen such that in each direction every domain has an integer
number of cells. But this would mean, in general, a different mesh
size for all subdomains, which may be far from the optimal
value. Therefore, it is advantageous (for large systems and moderate
polydispersities) to choose a global $a_x$ and $a_y$ instead, and
restrict the domain boundaries to this grid.

The domain decomposition method proposed in this paper is based on the
\emph{orthogonal recursive bisection} algorithm \cite{Salmon90} with
axis-aligned domain boundaries. The basis of the algorithm is the
hierarchical subdivision of the system. Each division represents
recursive halving of domains into two subsequent domains. The
advantage of such a division is an easy implementation of load
balancing, which can be realized at any level, simply by shifting one
boundary.

\begin{figure}[t]
\centering
\includegraphics*[width=0.65\textwidth]{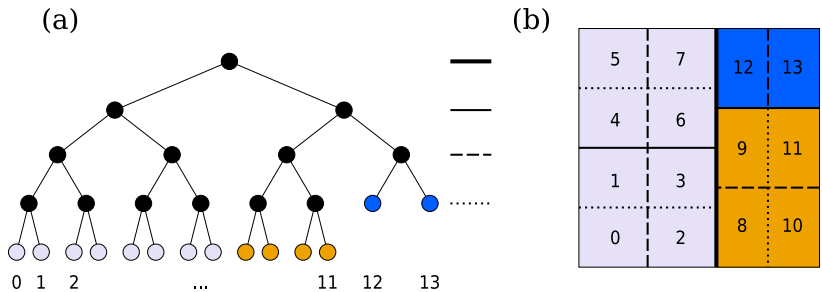}
\caption{(color online) An initial hierarchical decomposition of the simulation 
domain for $N_p=14$.}
  \label{Fig-8}
\end{figure}

First, we have to group the $N_p$ processors (where $N_p$ is not
required to be a power of two) hierarchically into pairs.  The
division algorithm we use is the following: We start at level $0$ with
one \emph{node}\footnote{These are abstract nodes in a tree rather
  than (compounds of) CPUs.}, which initially is a \emph{leaf} (a node
with no children) as well. A new level $l$ is created by branching
each node of level $l{-}1$ in succession into two nodes of level $l$,
creating $2^l$ leaves. This continues until $2^l<N_p\le
2^{l+1}$. Then, only $N_p-2^l$ leaves from level $l$ are branched from
left to right, cf.\ Fig.~\ref{Fig-8}(a).

Next, we have to assign a domain to each leaf/processor. In the
beginning, having no information about the system, all domains should
have the same size. Actually, their sizes equal only approximatively
due to grid restriction described above, cf.\
Fig.~\ref{Fig-9}(a). To achieve this, the recursive division
of the sample is done according to the tree just described. Each
non-leaf node represents a bisection with areas corresponding to the
number of leaves of its branches (subtrees). The direction of the cut
is always chosen as to minimize the boundary length.

The hierarchical decomposition method provides the possibility of
quick searches through the binary tree structure. For example, the
task to find the corresponding subdomain of each particle after load
balancing requires a search of order ${\cal O}(\log(N_p))$ for $N_p$
processors. With respect to bookkeeping overhead, a 
further advantage of this decomposition scheme is that local load 
imbalance does not necessarily affects higher level subdomain 
boundaries. For example, if particle exchange takes place across a 
low level domain boundary only this boundary will move leaving the 
others untouched.

\subsection{Adaptive load balancing}
\label{LoadBalancing}

For homogeneous quasi-static systems, the initially equal-sized
subdomains provide already a reasonably uniform load distribution, but
for any other case the domain boundaries should dynamically move
during the simulation. In the load balancing function, we take
advantage of the possibility provided by MPI to measure the wall clock
time accurately. For every time step, the processors measure the
computational time spent on calculations and broadcast it, so that all
processors can decide simultaneously whether or not the load balancing
procedure has to be executed. To quantify the global load imbalance,
the relative standard deviation of the elapsed CPU time in this time
step \footnote{Assuming exclusive access to the computing resources on
  every processor, we identify wall clock time and CPU time throughout
  this work.} is calculated via the dimensionless quantity
\begin{equation}
  \sigma_T \equiv \frac{1}{\langle T \rangle}
  \sqrt{{\langle T^2 \rangle} - {\langle T \rangle}^2},
  \label{cputime-standard-deviation}
\end{equation}
where the average is taken over the processors.

\begin{figure}[t]
  \centering
  \raisebox{32ex}{a)}\quad  \includegraphics*[width=0.3\textwidth]{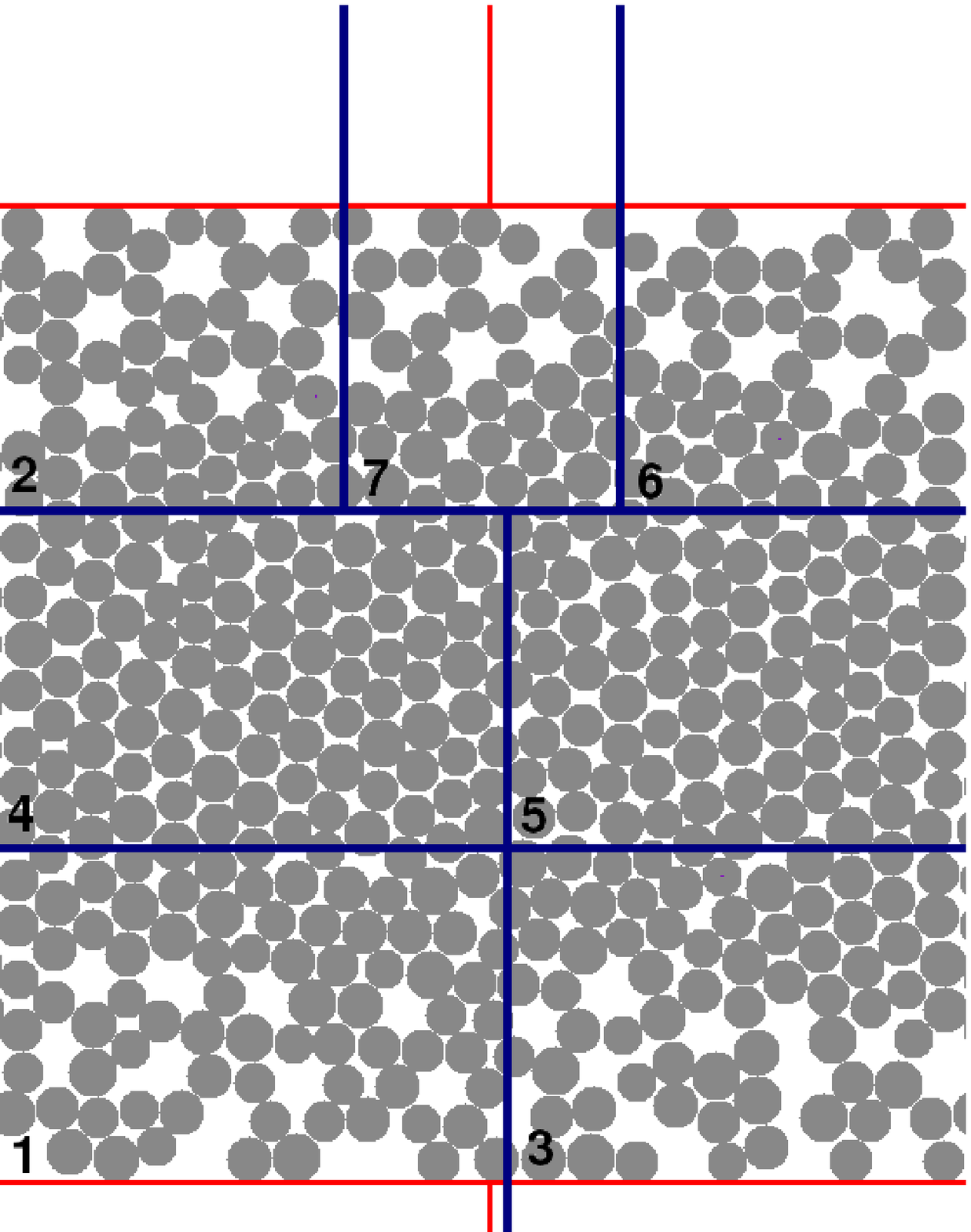}
\quad \raisebox{17ex}{$\Rightarrow$} \raisebox{0ex}{\includegraphics*[width=0.3\textwidth]{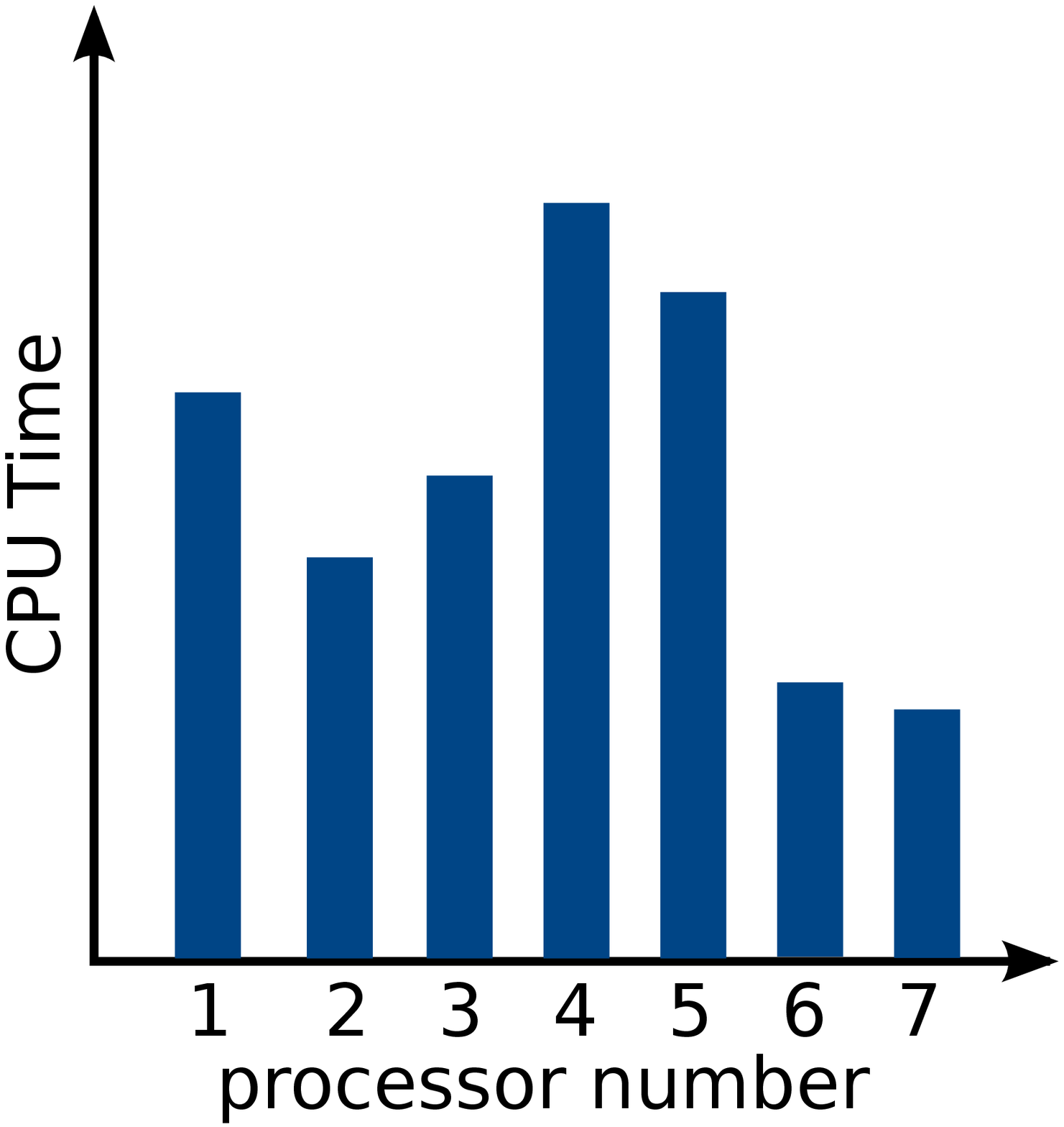}}\vspace{3ex}\\
  \raisebox{32ex}{b)}\quad  \includegraphics*[width=0.3\textwidth]{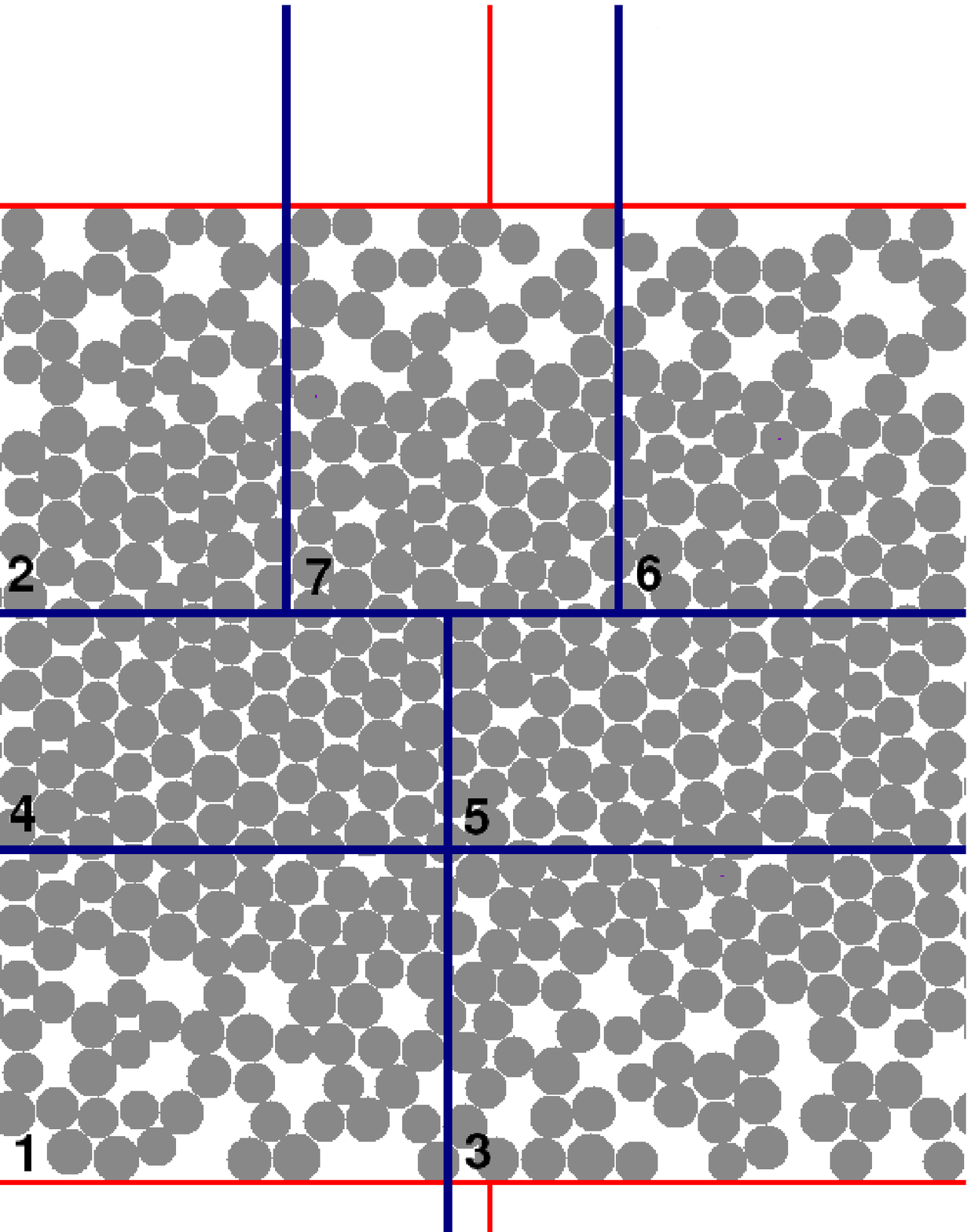}
\quad \raisebox{17ex}{$\Rightarrow$} \raisebox{0ex}{\includegraphics*[width=0.3\textwidth]{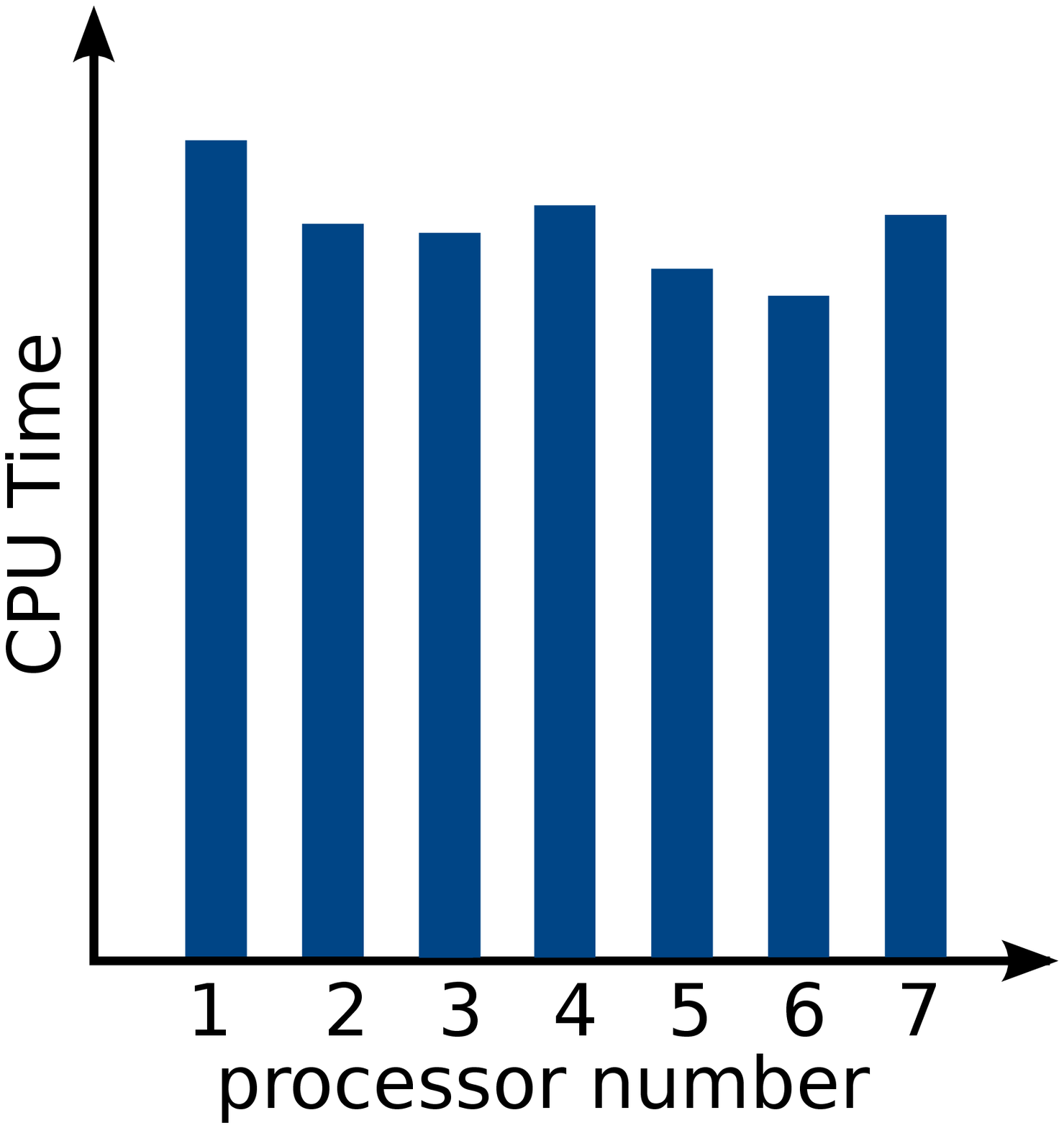}}
 \caption{(color online) (a) Geometrical domain decomposition at the
    beginning of the simulation leads to an unbalanced distribution
    of the load over the processors. (b) After load balancing, the volume
    of the subdomains belonging to different processors vary according to the
    CPU time it needed in the previous time step and the load
    distribution over the processors becomes more even.}
  \label{Fig-9}
\end{figure}

A threshold value $\sigma_T^*$ is defined to control the function of
the load balancing algorithm: If $\sigma_T<\sigma_T^*$, then the
simulation is continued with the same domain configuration, otherwise
load balancing must take place. This load balancing test is performed
by all processors simultaneously, since all of them have the
necessary data. The result being the same on all processors, no more
communication is needed.

If the above test indicates load imbalance, we have to move the domain
boundaries. This may happen at any non-leaf node of the domain hierarchy
tree. The relevant parameter for the domain division is the
\emph{calculating capacity} of the branches, which is defined as
\begin{equation}
\nu_j = \sum_i \frac{V_i}{T_i},
\end{equation}
where $T_i$ and $V_i$ are the CPU time and volume of domain $i$, 
respectively, and the summation includes all {\em leaves} under branch $j$. Let us
denote the two branches of a node as $j$ and $k$, then the domain must
be bisectioned according to
\begin{equation}
  \tilde \nu_j\equiv\frac{\nu_j}{\nu_j+\nu_k} \quad\text{and}
  \quad\tilde \nu_k\equiv1-\tilde \nu_j
  \quad.
\end{equation}
The above procedure is repeated for all parent nodes. If the
size of a domain was changed, then all subdomain walls must be
recalculated as even with perfect local load balance the orientation
of the domain boundary may be subject to change. Note that
boundaries must be aligned to the grid boundaries as explained in
Sec.~\ref{DomainDecompos}.

As an example, let us consider the situation of
Fig.~\ref{Fig-8} at the node of level $0$ with
branch $1$ to the left and branch $2$ to the right. If all $T_i$ would
be the same, then $\tilde\nu_1=8/14$ and $\tilde\nu_2=6/14$, just as 
the initial configuration. Let us now assume that the
processors 12 and 13 [top right in Fig.~\ref{Fig-8}(b)] are
only half as fast as the others, thus, the elapsed time is twice as
much. In this case $\tilde\nu_1=8/13$ and $\tilde\nu_2=5/13$, so the
thick, solid division line moves to the right. Furthermore, the thin,
solid division line on the right moves up from the position $4/6$ to
$4/5$.

Figure \ref{Fig-9} shows how load balancing improves the CPU
time distribution over seven processors. The initial geometrical
decomposition leads to an uneven workload distribution because of the
inhomogeneous density of the original particle configuration
[Fig.~\ref{Fig-9}(a)]. However, the load balancing function
manages to approximately equalize the CPU times in the next time step
by moving the borders [Fig.~\ref{Fig-9}(b)].

\section{Numerical results}
\label{NumericalResults}

In the following, we present the results of test simulations for
different systems performed by the parallel code. The main question to
answer is how efficient is the parallel code, i.e.\ how much could we
speed up the calculations by means of parallelization. The sensitivity of 
the performance to the load balancing threshold is also studied. The 
partially parallel updates at the domain boundaries is the main consequence 
of parallelization, which may make a difference in the results compared 
to the sequential implementation. Therefore, we investigate the impact of 
parallelization on the number of iterations and on the physical properties 
of the solutions.

\subsection{Performance of the force calculation}
\label{PerrrformanceForceCalc}

In this section, we test the efficiency of the parallel algorithm solely
with respect to the force calculation. In general, it is the most time
consuming part of the contact dynamics simulation (see
Sec.~\ref{CPUTimeAnalysis}), so the efficient parallelization of the
iteration scheme is necessary for the overall performance.

\begin{figure}[t]
\centering
\includegraphics*[width=0.35\textwidth]{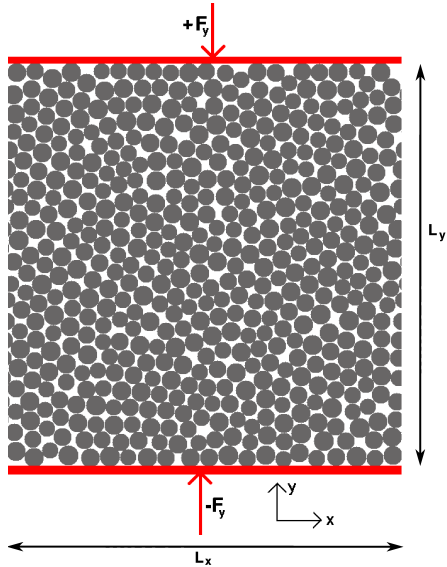}
\caption{(color online) The simulation setup used for performance tests. The system 
is confined by two lateral walls in $y$ direction (exerting a pressure 
of $0.25$ natural units), and periodic boundary conditions are applied 
in $x$ direction. The packings contain $500$, $8000$, and $10^6$ 
particles with $L_x{=}20$, $20$, $100$ and $L_y{=}20$, $320$, $10000$, 
respectively. The polydispersity in the small and medium systems amounts to
$20\%$, while the large system is monodisperse.}
\label{Fig-10}
\end{figure}

\begin{figure}[h]
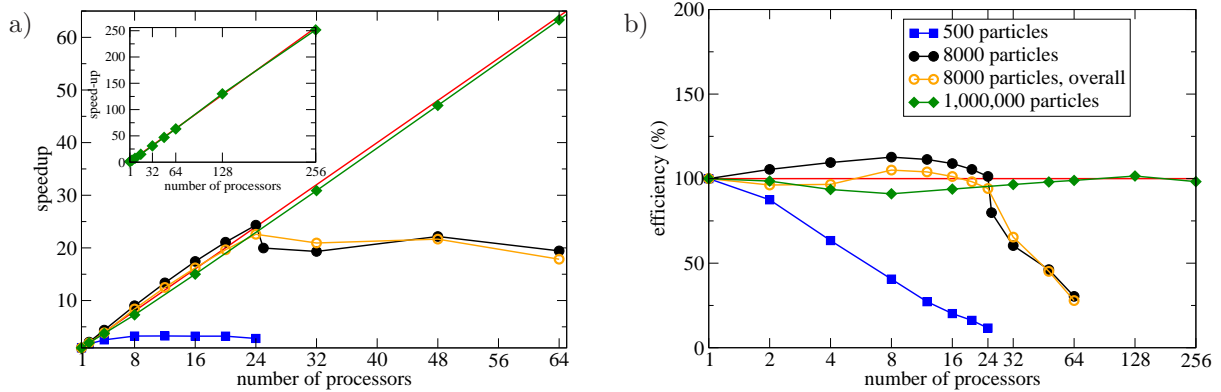

\centering
\raisebox{31.5ex}{a)}\includegraphics*[width=0.43\textwidth]
{Figure-11-a.eps}\qquad
\raisebox{31.5ex}{b)}\includegraphics*[width=0.45\textwidth]
{Figure-11-b.eps}
\caption{(color online) (a) Speedup and (b) efficiency of the force 
calculations for a small system with $500$ particles (full squares), 
a medium system with $8000$ particles (full circles), 
and a large system with $10^6$ particles (full diamonds). The open 
circles present the overall efficiency for the medium sized
system.}
\label{Fig-11}
\end{figure}

To focus just on the force calculation, we chose test systems where
large scale inhomogeneities are absent and adaptive load balancing is
unnecessary. Thus, dense static packings of $500$, $8000$, and $10^6$ 
particles with periodic boundary conditions in one direction and 
confining walls in the other were set up [see Fig.~\ref{Fig-10}].
The calculations started with no information about the contact forces 
and the simulation was stopped when the local convergence criterion is 
fulfilled (see Sec.~\ref{CD-algorithm}). Of course, this requires a 
different number of iterations depending on the system size and number 
of processors.
In order to get rid of perturbing factors like input/output
performance, we measured solely the CPU time spent in the iteration
loop. Figure~\ref{Fig-11} summarizes the test
results, which show that if the system is large compared to the
boundary regions, the efficiency is about $100\%$, which is equivalent
to a {\em linear} speedup. The smallest system is inapt for
parallelization, as already for only 4 processors the boundary regions
take up $20\%$ of the particles, which induces a large communication
overhead. The same fraction of boundary particles is 
reached around $N_p{=}32$ for the medium sized system with $8000$ 
particles. Therefore, one would expect the same performance for 
$N_p{=}4$ and $32$ for the small and medium sizes, respectively. 
In addition to the above mentioned effect, the efficiency of the 
medium system breaks down at $N_p{=}24$ due to special architecture 
of the distributed memory cluster used for simulations (Cray-XT6m 
with 24 cores per board), since the speed of the inter-board 
communications is much slower than the intra-board one. The observed 
efficiency values over $100\%$ are possible through caching, which was 
already observed in molecular dynamics \cite{LAMMPS}. The largest system 
has a large computation task compared to the boundary communication, which 
is manifested in almost $100\%$ efficiency. On the other hand, it is 
also too large for significant caching effects producing over $100\%$ 
efficiency. However, a gradual increase in the efficiency is observed 
as the domain size (per processor) decreases with increasing the number 
of processors.

For the medium sized system, we also measured the overall performance
including time stepping and load balancing. For this purpose, the
top wall was removed and the 
bottom wall was pushed upwards in order to generate internal dynamical 
processes, which unbalances the load distribution. As shown in
Fig.~\ref{Fig-11}, there is no significant
difference in efficiency due to the fact that time stepping and contact
detection are perfectly parallelizable processes.

\subsection{Load balancing threshold}
\label{Load-balancing-threshold}

In Sec.~\ref{LoadBalancing}, we defined the load balancing threshold 
$\sigma_T^*$ for the relative standard deviation of the elapsed CPU 
time on different processors, above which load balancing takes place. 
While the load balancing test is performed at each time step, the 
frequency of load redistribution is determined by the choice of 
$\sigma_T^*$. On the one hand, if the subdomain redivision happens 
frequently, a waste of CPU time is avoided because of even load 
distribution. On the other hand, the change of domain boundaries 
requires extra communication and administration. Doing this too 
often leads to unwanted overhead.

\begin{figure}[t]
\centering
\includegraphics*[width=0.45\textwidth]
{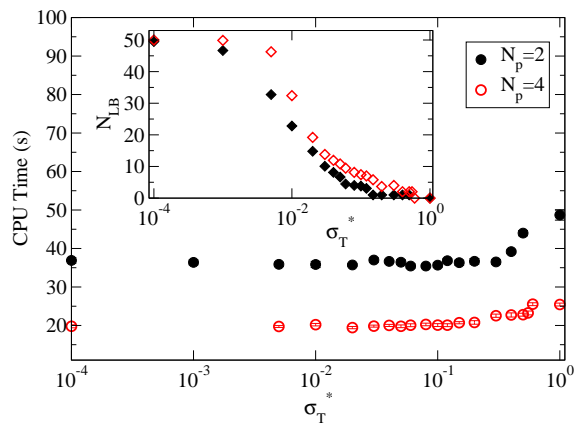}
\caption{(color online) CPU time as a function of the load balancing 
threshold $\sigma_T^*$. The simulation runs over $50$ time steps 
with $2$ or $4$ processors. The inset shows the number of load 
balancing events versus $\sigma_T^*$.}
\label{Fig-12}
\end{figure}

For load balancing, contact dynamics has the advantage, compared to
other DEM methods, that the configuration changes rather infrequently
(with respect to CPU time), because the force calculation with
typically $50{-}200$ iteration sweeps (for reasonably accurate 
precision of contact forces) dominates the computation. Thus, even 
taking the minimal value of $\sigma_T^*{=}0$ does not lead to 
measurable overhead. Moreover, in our implementation the domain 
boundaries must be on the cell grid, which avoids unnecessary small 
displacements of the domain walls. Hence, the optimal value of 
$\sigma_T^*$ is the minimal one as shown in 
Fig.~\ref{Fig-12}.

\subsection{Increase of the iteration number with the number of processors}
\label{Sec-NI-NP}

\begin{figure}[t]
\centering
\raisebox{36ex}{a)}\quad
\raisebox{18ex}{
\includegraphics*[width=0.33\textwidth]{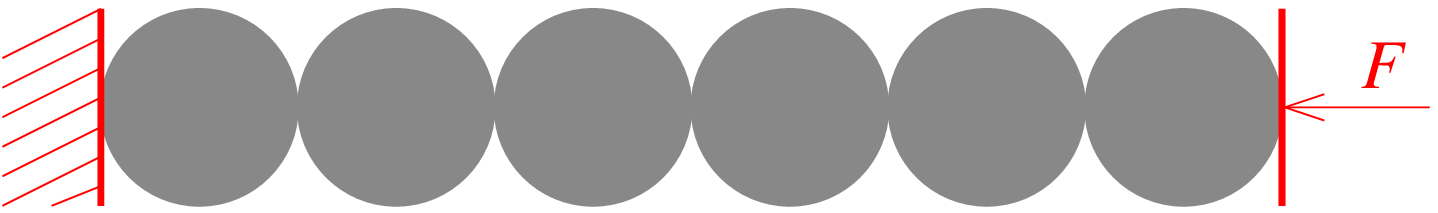}}\qquad
\raisebox{36ex}{b)}\quad
\includegraphics*[width=0.5\textwidth]{Figure-13-b.eps}
\linebreak
\raisebox{36ex}{c)}\quad\raisebox{3ex}{
\includegraphics*[width=0.33\textwidth]
{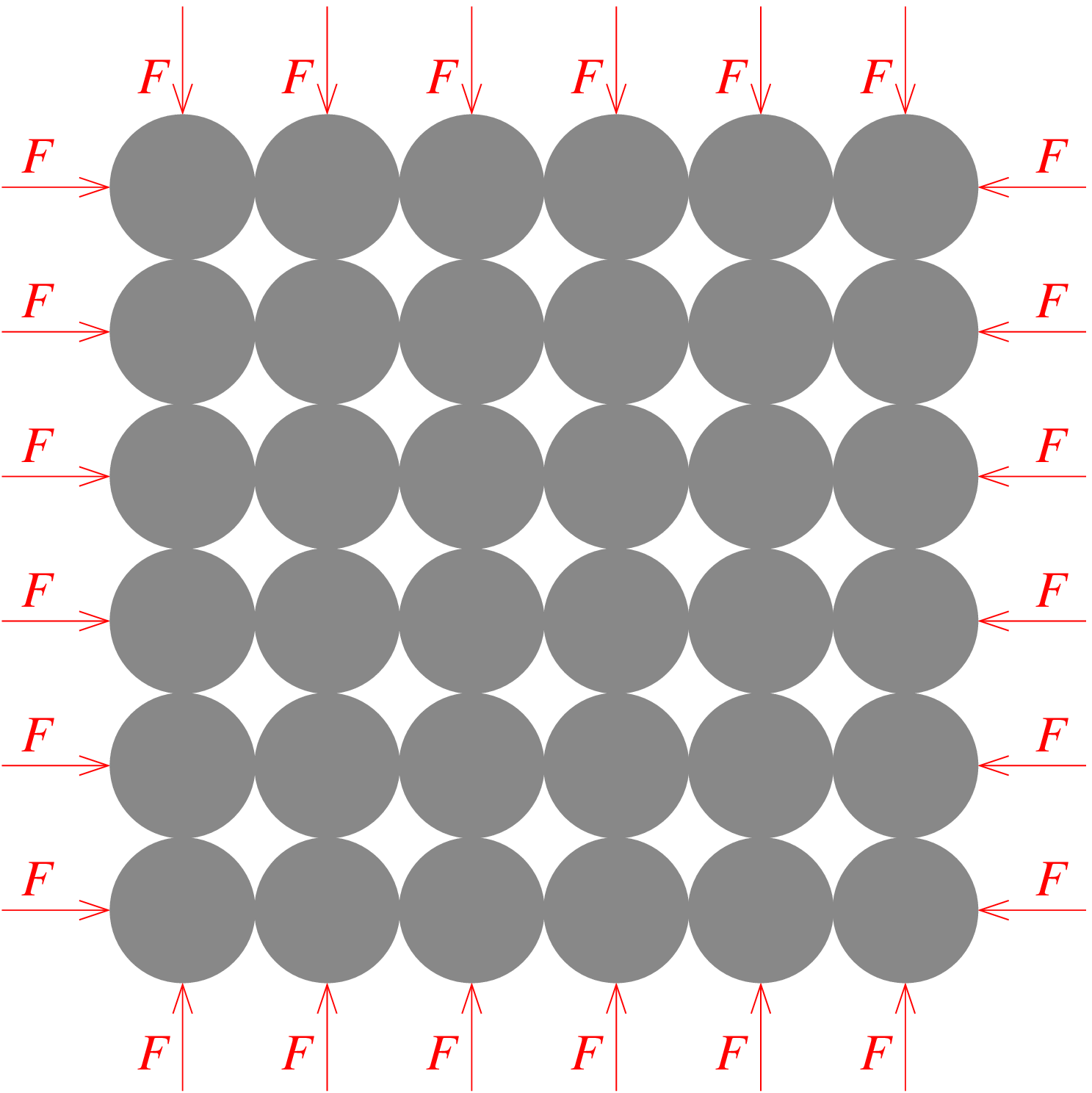}}\quad
\quad\raisebox{36ex}{d)}\quad\raisebox{0ex}
{\includegraphics*[width=0.5\textwidth]
{Figure-13-d.eps}}
\caption{(color online) (a) A chain of $n$ touching monodisperse 
particles, which are compressed with a force $F$. (b) The number 
of iterations needed to reach a given accuracy scaled by the 
value for a single processor ($\tilde N_I$) vs.\ the number 
of processors. The data points are simulation results, and 
the lines are linear fits (see text). (c) An ordered 
configuration of monodisperse particles, where the external 
forces $F$ push the outer particles inwards. (d) $\tilde N_I$ 
vs.\ $N_p$, where open circles denote the simulation results 
and the crosses are the theoretical estimations.}
\label{Fig-13}
\end{figure}

In the iteration scheme of contact dynamics, the forces relax towards
the solution in a diffusive way \cite{Unger02}. The diffusion constant
was found to be
\begin{equation}
\label{Eq:diffusion_constant}
D=q\frac{4\,r^2\,N_I}{\Delta t},
\end{equation}
where $\Delta t$ is the time step, $r$ is the diameter of a particle,
and $q$ is a constant depending on the update method: $q_p{=}0.5$ for
parallel and $q_s{\simeq}0.797$ for random sequential update. Thus the 
diffusion coefficient of the parallel update, $D_p$, is smaller than
that of the sequential update $D_s$, for a given set of parameters 
$N_I$, $\Delta t$, and $r$. Boundaries between sub-domains
handled by different processors behave like parallel update, since the
new information only arrives at the end of an iteration sweep. It is
therefore expected that the same system requires more iterations in 
the multiprocessor version, as the number of iterations is inversely 
proportional to the diffusion constant.

We test this conjecture on two examples: Let us first consider a linear chain
of $n$ touching identical particles placed between two perpendicular
plates [cf.\ Fig.~\ref{Fig-13}(a)]. We suddenly switch on a compressing
force on one side wall, while keeping the other wall fixed. The
resulting contact forces are calculated by the iterative solver. In
order to estimate the number of required iterations, we define the
effective diffusion coefficient as of \cite{Revathi93}:
\begin{equation} \label{Eq:Effectivediffusioncoefficient}
\overline D=D_pp+D_s(1-p),
\end{equation}
where $p$ is the portion of the chain with a parallel update. In
general, for each boundary one particle diameter is handled parallel
and the rest sequential, which gives $p{=}N_p/n$. This is compared
to the numerical results in Fig.~\ref{Fig-13}(b). While in principle
there is no fit parameter in 
Eq.~(\ref{Eq:Effectivediffusioncoefficient}), by adjusting the ratio
to $D_s/D_p{=}1.53$ we get an almost perfect agreement for all
different system sizes, as shown in Fig.~\ref{Fig-13}(b). This fitted
value is $4\%$ smaller than the theoretical estimation of
\cite{Unger02}.

We have tested this scenario in a similar two-dimensional setup, where 
the forces were directly applied to the boundary particles as shown in 
Fig.~\ref{Fig-13}(c). The number of iterations required for the 
prescribed force accuracy increases with the number of processors in a 
sub-linear manner [Fig.~\ref{Fig-13}(d)]. This is expected as the fraction 
of boundary particles in a two-dimensional system scales as 
$\sqrt{N_p/n}$. The theoretical estimation used in the above one
dimensional example with $D_s/D_p{=}1.53$ is in good agreement with
the results of the two dimensional system as well. The graph of
simulation results is characterized by plateaus (e.g.\ between 
$N_p{=}2{-}4$ and $6{-}8$), where the convergence rate is dominated 
by the higher number of domain walls in one direction.

Let us conclude here that the slower parallel diffusion part takes place in a
portion $p{\propto}\sqrt{N_p/n}$ of the two dimensional system, which is 
negligible in reasonably large systems. For example for the medium sized 
system of $8000$ particles, we get $p{\simeq}4\%$ for $N_p{=}16$, which would lead to about 
$2\%$ increase in the iteration number. The measured value was about
$1\%$ justifying the insignificance of the iteration number increase
in large systems. Indeed, we do not see a decrease in efficiency due
to an increase of the iteration number for large parallel systems in
Fig.~\ref{Fig-11}.

\subsection{Influence of the parallelization on the physical
  properties of the solutions}
\label{PhysicalProperties}

As a last check, we tested the physical properties of the system
calculated by different number of processors. It is known that in the
rigid limit, the force network of a given geometrical packing is not
unique \cite{Shaebani09-2,Unger05}. Running the contact dynamics with
different random seeds (for the random sequential update) leads to
different sets of contact forces, which all ensure the dynamical
equilibrium. The domain decomposition also changes the update order and
the solutions will be microscopically different. Thus, a direct
comparison is impossible and we have to resort to comparing
distributions.

\begin{figure}[t]
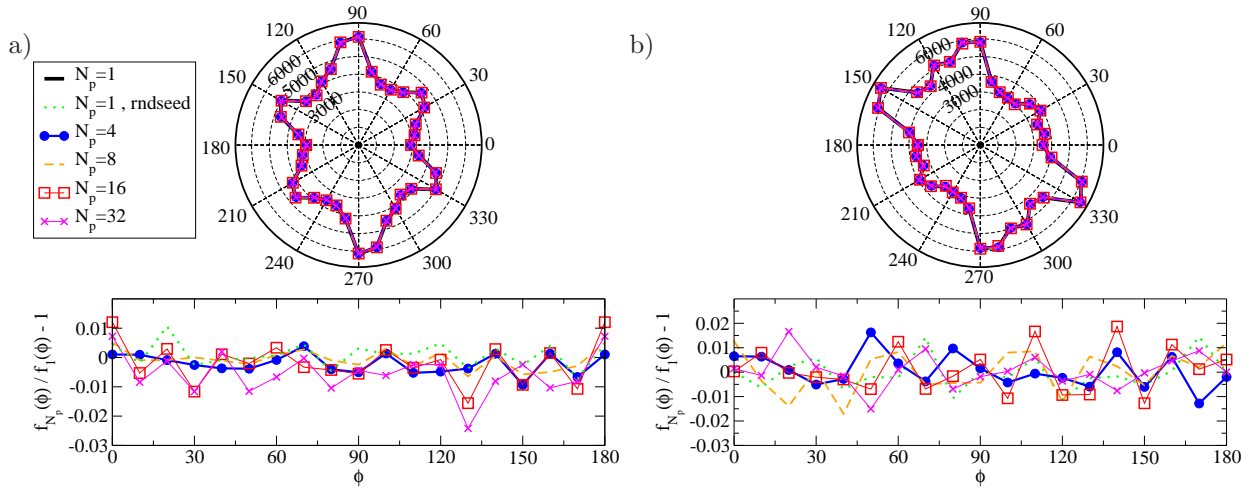

\centering
\raisebox{38ex}{a)}\includegraphics*[scale=0.30]{Figure-14-a.eps}
\raisebox{38ex}{b)}\includegraphics*[scale=0.30]{Figure-14-b.eps}
\caption{(color online) Angular distribution of the contact force 
orientations in (a) the relaxed static packing and (b) the sheared system 
with moving confining walls, with $8000$ frictional particles calculated 
for different number of processors.}
  \label{Fig-14}
\end{figure}

We first investigate the distribution of the contact force
orientations $f(\phi)$ in the relaxed system of $8000$ particles
described in Sec.~\ref{PerrrformanceForceCalc}. The contact forces are
calculated from scratch for the given geometry and boundary conditions
using different number of processors. Since the system is very
tall ($L_y/L_x{=}16$), it is divided only vertically for up to $N_p{=}16$,
while for $N_p{=}32$ the 16 domains are cut horizontally as well. 
The orientation of each contact force is defined as 
$\phi{=}\arctan(R_y/R_x)$. The distributions of the contact force 
orientations, $f_{N_{_p}}(\phi)$, are compared for several values 
of $N_p$ in Fig.~\ref{Fig-14}(a). The range of possible values for 
$\phi$ ($[0,\pi]$) is divided into $18$ bins, and each data point 
in the figure corresponds to total number of contacts in the 
same bin. For comparison, we have presented the 
results of the simulations with $N_p{=}1$ for two different random seeds 
as well. The match among the different runs are so good that the 
curves coincide. Hence, we also plot the relative difference 
$f_{N_{_p}}(\phi){/}f_{1}(\phi){-1}$ to the non-parallel run for 
comparison, which shows negligible random noise. Evidently, 
parallelization has no systematic impact on the angular distribution 
of the contact forces. Similar results were obtained when the system is sheared by the 
horizontal confining walls moving with a constant velocity in 
opposite directions as shown in Fig.~\ref{Fig-14}(b). 

\begin{figure}[t]
\centering
\includegraphics*[width=0.85\textwidth]
{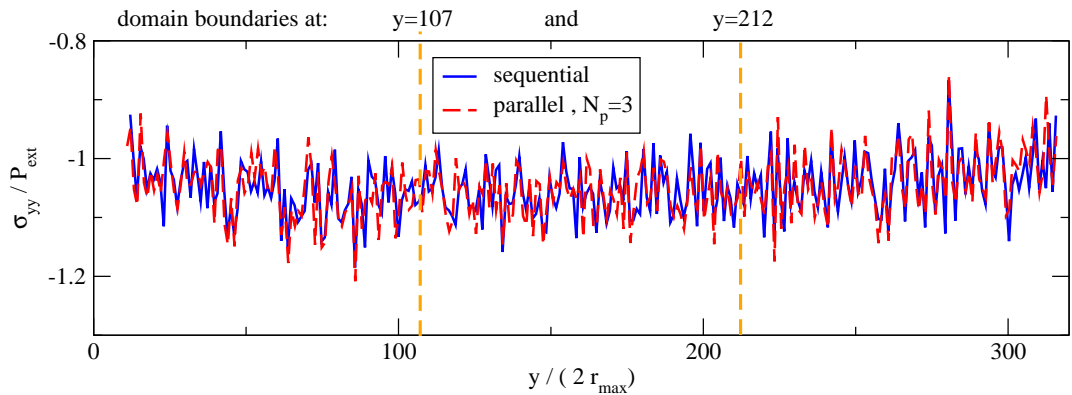}
\caption{(color online) $\sigma_{yy}(y)$ scaled by the external 
pressure $P_\text{ext}$ in terms of the height $y$ scaled by the 
diameter of the largest particle in the system ($2\,r_\text{max}$). 
The results obtained by the non-parallel code are compared with 
those obtained by the parallel code for $N_p=3$.}
\label{Fig-15}
\end{figure}

We also calculate the $\sigma_{yy}$ component of the stress tensor 
as a function of the distance $y$ from the bottom wall in the same
system. $\sigma_{yy}(y)$ at a given height $y$ is averaged over a 
horizontal stripe of width $dy{=}2r_\text{max}$, where 
$r_\text{max}$ is the largest particle radius in the system.
The system height is thus divided into nearly $320$ stripes.
Figure \ref{Fig-15} displays the results obtained by the 
non-parallel code as well as the parallel code with $N_p{=}3$. 
In the parallel case, the system is divided horizontally 
into three parts. The results of the parallel run match 
perfectly with the one of the non-parallel run. Especially, 
no kind of discontinuity or anomaly is observed at 
$y \simeq 107$ and $y \simeq 212$, where the interfaces 
between the processors are located.

\section{Conclusion and Discussion}
\label{conclusions}

We have presented an efficient parallel version of a contact dynamics
method in this work, which allows for large-scale granular simulations 
with almost $100\%$ efficiency. We aimed at the full parallelization 
of the code with hierarchical domain decomposition and dynamic load 
balancing, in which the interface area between subdomains is also minimized. 
The parallel code is hence applicable to a broad range of densities and 
different simulation conditions.

The force calculation in CD is done by an iterative scheme, which
shows an instability if more than about half of the contacts are calculated 
in parallel. The iteration scheme was kept domain-wise sequential while
data across the domain boundaries is exchanged after each iteration sweep,
ensuring that the iteration is stable for all system sizes. 
It is known that the CD iterative scheme approaches the solution in 
a diffusive manner. The diffusion constant is smaller for parallel 
update, which happens at domain boundaries. However, this overhead 
is proportional to the square root of the number of processors  
divided by the number of particles (in 2D), which vanishes for large 
systems. Regarding this as the only impact of the 
parallelization on the convergence, it must be expected that the 
efficiency is not affected by modifications at the local level 
i.e.\ non-spherical particles, three-dimensional particles, more 
sophisticated contact laws, etc. Of course, those can deteriorate 
the convergence per se but the parallel version will simply 
``inherit'' that.

The other point of discussion raised here concerns the choice of the
mesh size and adjusting the subdomain borders to it. Communication
overhead was reduced because between iteration steps not all boundary
information is sent but only the relevant part of it. The subdomain 
wall position is only important if the particle size is not small 
compared to the system size. For large scale parallel applications 
this can only be a problem for highly polydisperse systems, for which 
the cell method for contact detection breaks down anyway.

The load balancing is done only at the end of each time step. Our investigations show that this happens 
rarely enough that load balancing overhead and CPU time fluctuations 
are negligible but often enough to achieve fast load balance. We used 
a global criterion for stopping the iteration scheme. This ensures 
that the physical properties of the tested samples do not show any 
difference compared to the non-parallel version of the code.

Blocking point-to-point communications were used to transfer data
among processors. Since our algorithm needs synchronization after
each iteration, non-blocking data transfer would not be advantageous.
The whole amount of data is transmitted in one single packet, which
reduces communication overhead over the pure data. This method
introduces parallel contact update at domain boundaries, which induces
an iteration number overhead due to the lower diffusivity of the
information in parallel update. This overhead vanishes, e.g. with 
the square root of the processor number over particle number in two 
dimensions, which is in general negligible.

An alternative method would be to use non-blocking communications for 
the iteration scheme, namely to immediately send a freshly updated 
contact force in the vicinity of the borders to the corresponding 
processors, while on the other side this would trigger an interrupt 
when the other processor immediately updates the received contact data. 
This prevents the mixture of sequential and parallel update schemes. 
However, we do not expect that the performance of the method is greatly 
enhanced by the use of non-blocking communication because the information 
of each contact force is sent individually and the overhead associated 
with the increase of the inter-processor communications significantly 
affects the performance.

The last point to discuss concerns the load balancing method. The most
exact method would be to consider the number of particles and/or
contacts in each subdomain to calculate their new
boundaries. Practically, this would cause difficulties, since each
processor is just aware of particles and contacts within its own
borders. The amount of calculations and communications between
neighboring processors to place the interface according to the current
contact and particle positions would make the load balancing a
computationally expensive process. This lead us to balance the load
further by dividing the simulation domain according to the current
subdomain volumes (not always proportional to the number of particles
and/or contacts), which is in fact a control loop with the inherent
problems of under- and over-damping.

\section*{Acknowledgments}
We would like to thank M.\ Magiera, M.\ Gruner and A.\ Hucht for 
technical support and useful discussions, and M.\ Vennemann for 
comments on the manuscript. Computation time provided by 
John-von-Neumann Institute of Computing (NIC) in J\"ulich is 
gratefully acknowledged. This research was supported by DFG Grant 
No.\ Wo577/8 within the priority program ``Particles in Contact''.

\end{document}